# ELECTRON PARAMETERS IN H2:Ar:CH4 PLASMA MAINTAINED BY MICROWAVE AND LASER RADIATION

S.V. Avtaeva[1*], A.E. Medvedev[1]

[1] *Institute of Laser Physics, Siberian Branch of the Russian Academy of Sciences, 15B,Lavrentyev Ave., Novosibirsk, 630090, Russia*

**s_avtaeva@mail.ru*

**Abstract.** Electron parameters in an atmospheric-pressure hybrid plasma in the $95H_2$:3Ar:$2CH_4$ mixture maintained by pulsed-periodic microwave radiation (2.47 GHz) and $CO_2$ laser radiation (10.6 μm) were studied**.** The plasma was generated in a quasi-cylindrical reactor, which was a resonator for $TM_{011}$-mode microwave radiation. It was shown that in the hybrid plasma emission spectra the $H_\alpha$ and $H_\beta$ lines are strongly broadened and can be approximated by the sum of two Lorentzian contours (dual-Lorentzian approximation). The Stark broadening of the $H_\alpha$ and $H_\beta$ lines was used to measure the electron density in the plasma, and the ratio of their intensities allowed to estimate the electron temperature. To simplify the description of the hybrid plasma properties the concept of two plasma states caused by the pulsed injection of laser energy into the plasma was introduced. The hybrid plasma state, corresponding to the plasma state during the laser pulse, is characterized by a high electron density of $(1\text{-}2.5)\cdot10^{17}$ $cm^{-3}$ and an electron temperature of ~0.5-0.6 eV. The electron density in the state corresponding to the time intervals between laser pulses ($1.8\cdot10^{14}$-$5\cdot10^{15}$ $cm^{-3}$) is more than an order of magnitude lower than the electron density in the first state, but it is 2-10 times higher than the electron density in microwave plasma ($(1\text{-}5)\cdot10^{14}$ $cm^{-3}$). The electron temperature in this state is 0.16-0.24 eV. The radius of the dense plasma region with an electron density of ~$10^{17}$ $cm^{-3}$ is estimated to be ~0.5-1 mm.

**Keywords:** hybrid plasma, microwave radiation, laser radiation, optical emission spectroscopy, Stark broadening, electron density, electron temperature.

## Introduction

Diamond-like polycrystalline films possess a number of beneficial properties, such as exceptional hardness and wear resistance, a low coefficient of friction, chemical inertness, biocompatibility, and optical transparency [1, 2]. These properties enable a wide range of applications for diamond-like films in fields including mechanical engineering, transport, medicine, optics, and the electronics industry. The synthesis of diamond-like coatings (DLC) from the gas phase using microwave (mw) plasma has become widely adopted. The high uniformity and stability of a non-equilibrium mw discharge, maintained at pressures significantly below atmospheric, allow for the production of high-quality cotings [2–4]. However, the typical growth rates for polycrystalline diamond coatings in such mw plasma generally do not exceed ten microns per hour [5–8]. These limitations on the growth rate of diamond-like films are due to the relatively low density of charged particles in low-and medium-pressure mw plasma, which in turn results in a low production rate of the chemically active plasma species required for coating synthesis [6–8].

A significantly higher density of charged particles, which could facilitate a faster growth rate for a synthesized layer, can be achieved in laser-produced plasma [9]. At atmospheric pressures the plasma generated by a $CO_2$ laser typically exhibits electron densities in the range of $10^{15}$–$10^{17}$ $cm^{-3}$ and electron temperatures of 1–2 eV [10–12]. However, laser-produced plasma is characterized by significant non-uniformity and a small surface area available for synthesis [13]. Furthermore, such plasma generally has a high gas temperature, approaching the electron temperature, whereas the synthesis of diamond-like films requires the substrate temperature to be maintained within the 800–1000 K range [14, 15].

Combining the advantageous properties of these two plasma sources could prove promising for increasing the deposition rate of diamond-like films while maintaining high coating quality. By simultaneously introducing mw radiation (2.45 GHz) and pulsed $CO_2$ laser radiation (10.6 μm) into a gas at atmospheric pressure, one can expect a substantial increase in the density of charged particles (compared to their density in a mw-generated plasma) along with a significant reduction in the non-uniformity typically associated with laser-produced plasma. The possibility of obtaining, within such a hybrid plasma, the plasma parameters required for the synthesis of high-quality DLCs warrants careful investigation.

In this work, optical emission spectroscopy (OES) is used to study the electron density and temperature in the $95H_2:3Ar:2CH_4$ hybrid plasma at atmospheric pressure sustained by microwave (2.47 GHz) and $CO_2$ laser (10.6 μm) radiation.

## Experimental setup

A dense carbon-containing plasma was generated within a reactor chamber which is a quasi-cylindrical mw cavity designed on the $TM_{011}$ mode [16]. The mw cavity consists of two cones (a small one *1* and a large one *2)* connected by a threaded joint *3*, which allows for adjustment of the cavity volume to achieve resonance with the mw generator (Fig. 1).

A gas mixture is fed into the reactor chamber through inlet *4*, creating an overpressure slightly above atmospheric (~0.1 atm). The composition of the gas mixture is regulated by a four-channel gas mixer, which controls the flow rate in each channel. When mw radiation is introduced into the resonator via input antenna *6*, a gas breakdown occurs in the region of the electromagnetic standing wave primary maximum, near the exit nozzle in the chamber's top cover, resulting in the formation of a mw plasma *7* with a characteristic dimension of approximately 20 mm. Pulsed-periodic radiation from a $CO_2$ laser *8* is directed by a copper mirror *9*, through an evanescent (cut-off) waveguide *10*, toward a lens *11* (ZnSe, *d* = 50 mm, *F* = 250 mm), which focuses the beam into the center of the mw plasma clot, thereby creating a hybrid plasma *12*. The diameter of the laser's focal spot is approximately 100 μm. As the $CO_2$ laser radiation enters the mw plasma clot, a bright flash of light is visible through the reactor window. A jet of plasma products *13* then exits the reactor through a cylindrical nozzle *14* with diameter of 1 mm and spreads toward a substrate *15*.

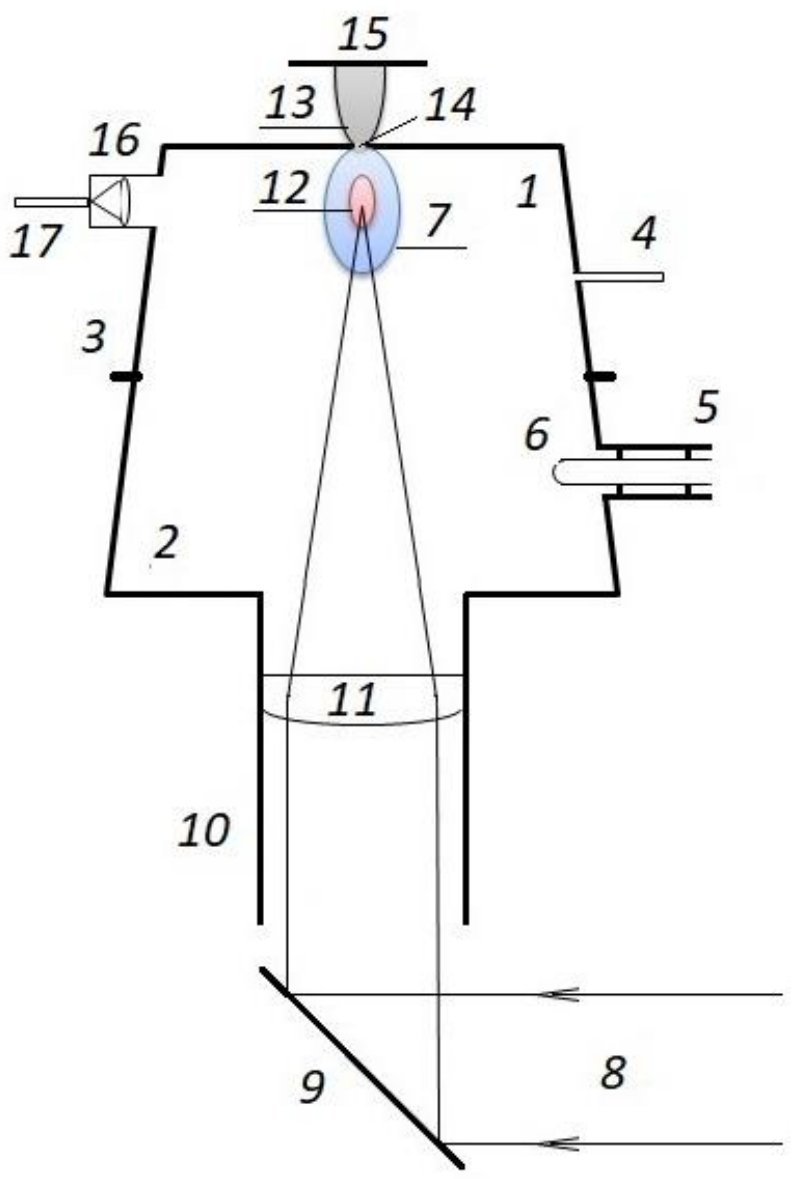

Figure 1. Schematic of hybrid plasma formation in the resonant cavity of the experimental reactor. 1, 2 – small and large reactor cones; 3 – junction area of the reactor cones; 4 – gas inlet; 5, 6 – mw cable and antenna for injection of mw radiation into the cavity; 7 – mw plasma plume; 8 – $CO_2$ laser beam; 9 – copper mirror; 10 – cutoff waveguide; 11 – long-focus lens; 12 – hybrid plasma plume; 13 – jet of hybrid plasma products; 14 – cylindrical nozzle; 15 – substrate; 16 – collimator for coupling radiation into the optical fiber 17 and further to the HDSA spectrum analyzer.

Plasma emission was collected by a collimator *16* with an input aperture 4 mm in diameter, directed into an optical fiber *17*, and subsequently fed into an HDSA spectrum analyzer (Ångstrom, Novosibirsk, Russia & High Finesse, Tübingen, Germany) with a resolution of 0.2–0.3 Å, depending on the wavelength (350–1000 nm). The acquisition time for the plasma spectra ranged from 2 to 30 s.

The plasma was ignited in a gas mixture of $95H_2$:3Ar:$2CH_4$ at a pressure slightly above atmospheric. The mw power supplied to the discharge was 570 W, with a pulse repetition frequency of 10 kHz and a duty cycle of 5. The frequencies $\nu_{LP}$ and the corresponding radiation power $\bar{P}_{LP}$ of the laser pulses, which had a duration of ~0.2–0.5 μs, are listed in Table. Also provided in the table are the pulse energy values $E_{LP}$, calculated using relation (1), as well as the ratio of the laser pulse frequency $\nu_{LP}$ to the mw pulse frequency $\nu_{mwP}$.

$$\bar{P}_{LP} = E_{LP} \cdot \nu_{LP}. \qquad (1)$$

Table. Parameters of $CO_2$ laser radiation pulses (λ=10.6 μm, τ≈0.2-0.5 μs).

| Pulse repetition frequency, kHz | Average power, W | Pulse energy, mJ | $\nu_{LP}/\nu_{mwP}$ |
|---|---|---|---|
| 3 | 170 | 56,7 | 0,33 |
| 15 | 500 | 33,3 | 1,5 |
| 30 | 700 | 23,3 | 3 |
| 80 | 700 | 8,7 | 8 |

The dependence of the average output power and pulse energy of a $CO_2$ laser on the pulse repetition frequency is shown in Fig. 2. As can be seen from the figure, an increase in the $CO_2$ laser pulse repetition frequency leads to an increase in the average laser output power, while the energy of the laser pulses decreases significantly.

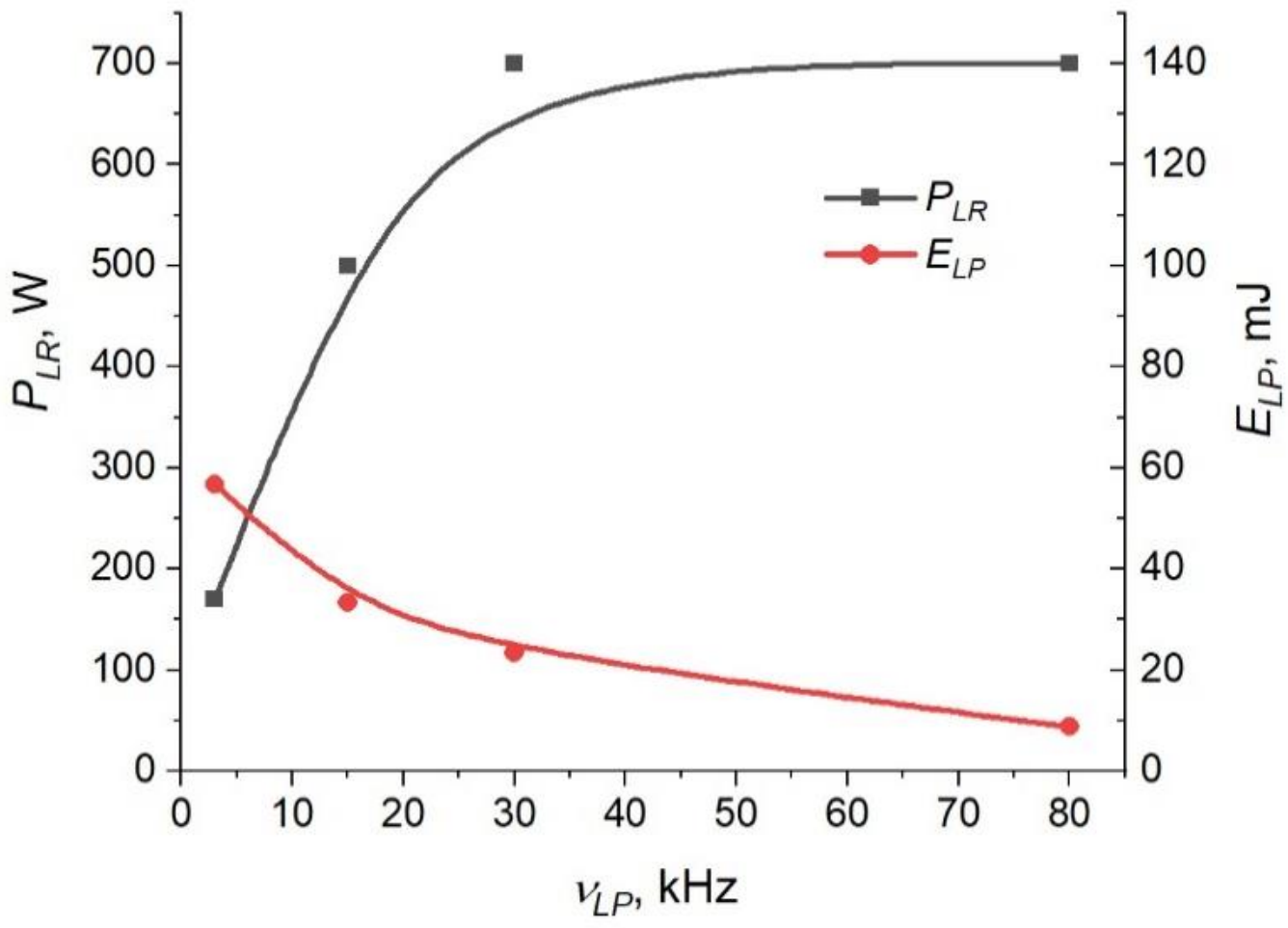

Fig. 2. Dependences of the average power and pulse energy of a $CO_2$ laser on the pulse repetition frequency.

## Results and Discussion

Typical spectra of mw and hybrid plasmas in the $95H_2$:3Ar:$2CH_4$ gas mixture are shown in Fig. 3. The mw plasma emission spectrum exhibits spectral lines of atomic hydrogen (Balmer series) and Swan system bands of molecular carbon ($C_2$). In the hybrid plasma emission spectrum the Swan system bands are very weak. Emission lines of argon atoms and ions are virtually absent in the spectra of both the mw and hybrid plasmas; this is due to the low proportion of argon in the gas mixture and the relatively low electron temperature.

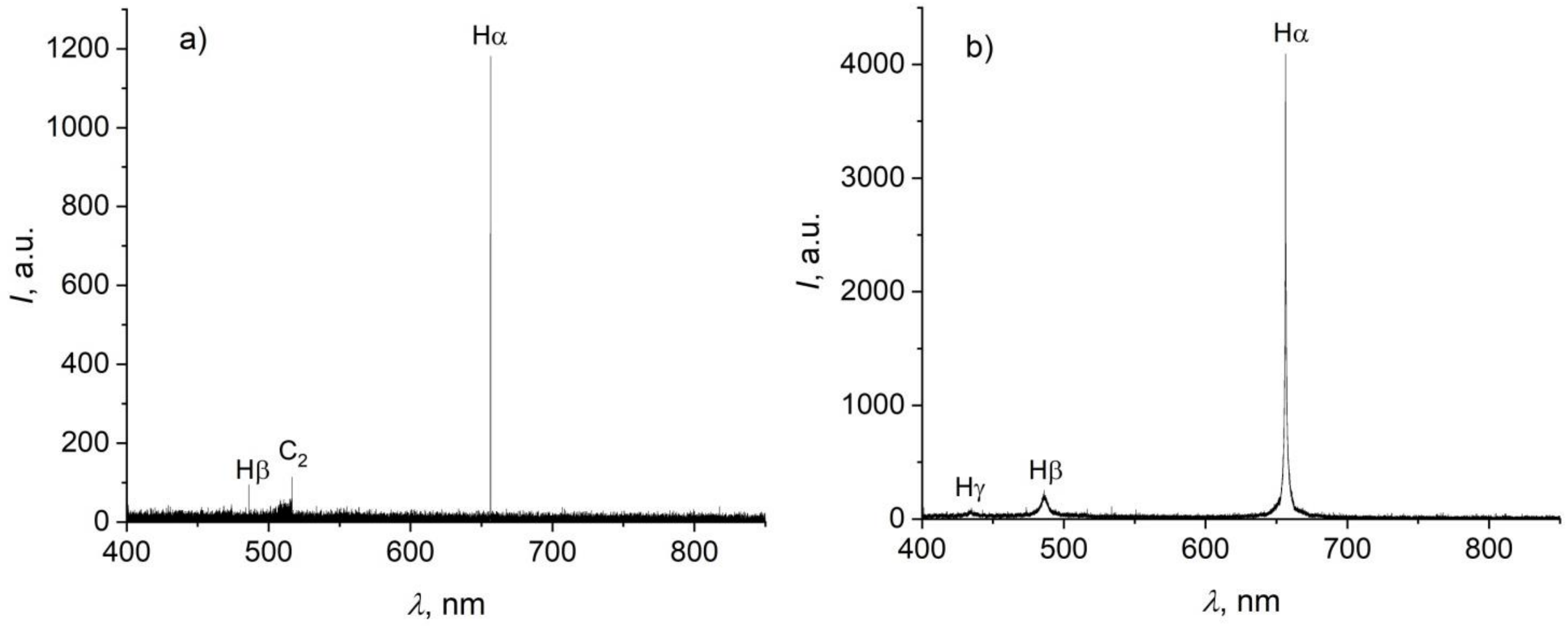

Fig. 3. Typical emission spectra of mw plasma (a) and hybrid plasma (b) in the $95H_2$:3Ar:$2CH_4$ mixture.

Atomic hydrogen lines in the spectrum of the hybrid plasma are significantly broadened compared to the corresponding lines in the spectrum of the mw plasma. Their width depends on the laser pulse frequency, the average laser power, and the laser pulse energy – parameters related by equation (1). The broadest lines are observed at laser frequencies of 15 and 30 kHz (Fig. 4a), conditions under which both the laser power and the laser pulse energy are relatively high. The experimentally observed $H_\alpha$ and $H_\beta$ lines profiles reflect the integrated emission characteristics of the hybrid plasma, as the time required to record these spectral line profiles is much longer than the characteristic timescales of the processes occurring within the plasma.

An estimation of the contributions from various broadening mechanisms for spectral line of Balmer series ($H_\alpha$, $H_\beta$) of hydrogen atoms in mw and hybrid atmospheric-pressure plasmas has shown that, at electron density exceeding $10^{14}$ cm$^{-3}$, Stark broadening is the primary mechanism of these lines broadening. An analysis of the $H_\alpha$ and $H_\beta$ line shapes emitted by the hybrid plasma revealed that they are well approximated by a sum of two Lorentzian functions (a dual-Lorentzian approximation) and cannot be accurately described by either a single Lorentzian function or a Voigt function. A typical example of the dual-Lorentzian approximation for the $H_\alpha$ line profile emitted by the $95H_2$:3Ar:$2CH_4$ hybrid plasma is presented in Fig. 4b.

Assuming that the two contours of the dual-Lorentzian approximation describe the Stark broadening of atomic hydrogen lines under the influence of microfields from two distinct plasma states, which differ in their charged-particle density, the half-widths of each contour were measured; subsequently, after isolating the Stark contribution, the electron densities corresponding to these plasma states were estimated.

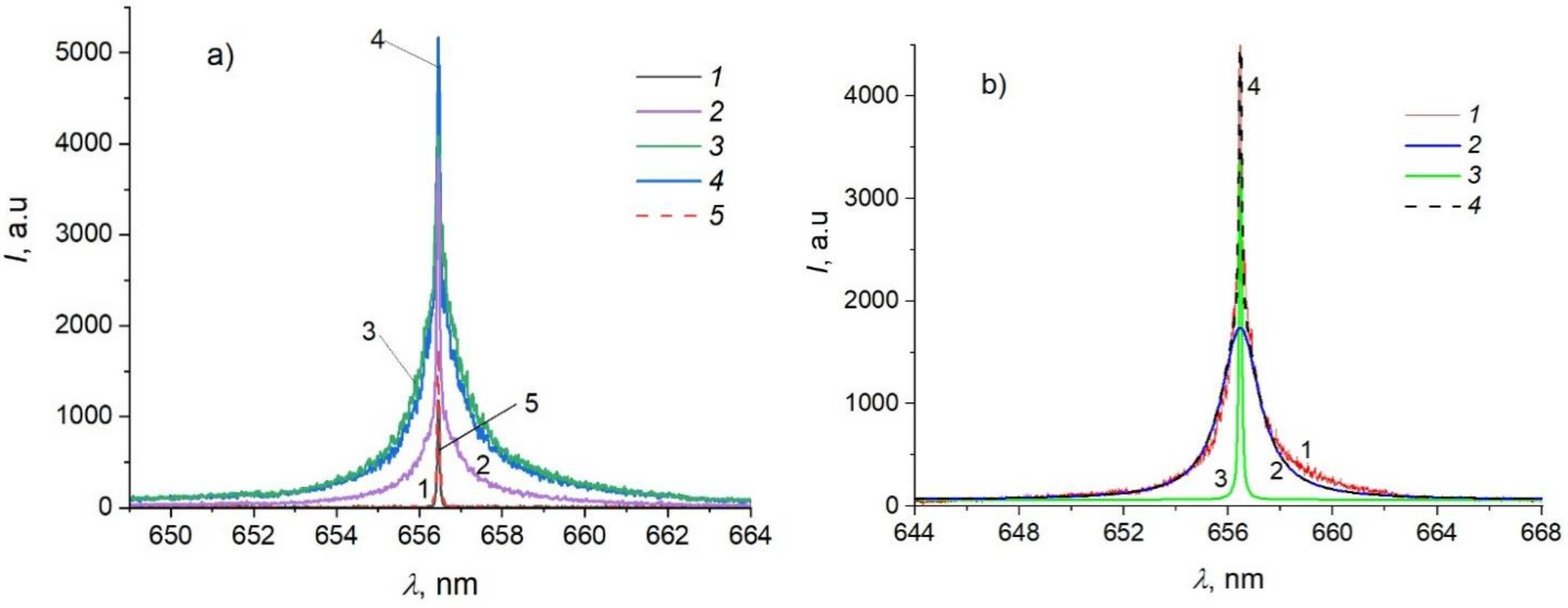

Fig. 4. a) $H_\alpha$ line shape in the emission spectra of the hybrid plasma in the $95H_2{:}3Ar{:}2CH_4$ mixture, as a function of laser pulse frequency; *1* – mw plasma, *2* – 3 kHz, *3* – 15 kHz, *4*- 30 kHz, *5* – 80 kHz; b) Dual-Lorentzian Lorentzian approximation of the $H_\alpha$ line profile ($CO_2$ laser pulse repetition rate is 30 kHz); *1* – experimental profile, *2* and *3* – broad and narrow contours of the dual-Lorentzian approximation, *4* – resulting profile of the approximation.

The instrumental function of the recording spectral system was measured using a helium-neon laser. Analysis of the instrumental profile showed that it has a Gaussian shape with a full width at half maximum of $\Delta\lambda_a$=0.02 nm. Doppler broadening of $H_\alpha$ and $H_\beta$ lines ($\Delta\lambda = 7.16 \cdot 10^{-7}\lambda\sqrt{T/\mu}$) was calculated for a gas temperature of 3000-3500K, which was determined from the intensity distribution of rotational lines in the (0–0) band of the Swan system of $C_2$ molecules ($d^3\Pi_g \rightarrow a^3\Pi_u$ - transition). The half-width of the resulting Gaussian broadening (instrumental plus Doppler) for atomic hydrogen lines at a temperature of 3500 K is ~0.034 nm for the $H_\alpha$ line and 0.029 nm for the $H_\beta$ line.

The Stark contribution to the broadening of the narrow component of the dual-Lorentzian approximation for the $H_\alpha$ and $H_\beta$ line profiles was extracted from the experimental profile using the formula [17]:

$$w_V = \frac{w_L}{2} + \left(\left(\frac{w_L}{2}\right)^2 + w_G^2\right)^{1/2} \qquad (2)$$

where $w_V$ is full width of the experimental observed shape, $w_L$ is Lorentz half-width due to the Stark effect, and $w_G$ is the half-width of the Gaussian profile resulting from instrumental and Doppler broadening. The broad component of the dual-Lorentzian approximation for the $H_\alpha$ and $H_\beta$ lines is significantly wider than the Gaussian profile caused by Doppler and instrumental broadening; therefore, the contribution of Gaussian broadening to this component could be neglected.

The conversion from the half-width of the Stark broadening of the $H_\alpha$ line to the electron density was performed using the formula [17]:

$$N_{e,\alpha}[\mathrm{m}^{-3}] = (w_S[\mathrm{nm}]\,/1.098\,)^{1.47135}\times 10^{23}, \qquad (3)$$

where $w_S$ is the full width at half maximum (FWHM) of the Stark contribution to the experimental line shape. The conversion from the half-width of the Stark broadening of the $H_\beta$ line to the electron density was performed using the formula [17]:

$$N_{e,b}[\mathrm{m}^{-3}] = (w_S[\mathrm{nm}]\,/0.94666\,)^{1.49}\times 10^{22}. \qquad (4)$$

Assuming a Boltzmann distribution of hydrogen atoms across energy levels, which is justified for a thermally equilibrium plasma, we also estimated the electron temperature corresponding to each of the two states of the hybrid plasma. In calculating the electron temperature, it was assumed that the relative intensity ($I$) of the $H_\alpha$ and $H_\beta$ lines for each plasma state is proportional to the area ($S$) of the corresponding component in the dual-Lorentzian approximation $I^j_{H\alpha}\sim S^j_{H\alpha}$ and $I^j_{H\beta}\sim S^j_{H\beta}$($j$=1, 2; (1 corresponds to the broad shape, 2 - to the narrow shape). The electron temperature $T_e$ was determined from the relation [18]

$$kT_e = \frac{E_{n=4}-E_{n=3}}{Ln\left(\frac{I^j_{H\alpha}\lambda^{H\alpha}_{ki}A^{H\beta}_{ki}\,g_{n=4}}{I^j_{H\beta}\lambda^{H\beta}_{ki}A^{H\alpha}_{ki}\,g_{n=3}}\right)}, \qquad (5)$$

where $\lambda_{ki}$ is wavelength of the spectral line, $A_{ki}$ is transition probability, $g_n$ is statistical weight of an energy level with quantum number $n$, $E_n$ is level energy.

Figure 5a shows electron densities calculated from the half-width of each component in the dual-Lorentzian approximation of the $H_\alpha$ and $H_\beta$ line profiles observed in the emission spectra of the hybrid plasma. As can be seen from the figure, the electron density derived from the broad component of the dual-Lorentzian approximation is in the range of $(1–2.5)\times10^{17}$ cm$^{-3}$; this likely corresponds to the high electron densities present in the hybrid plasma during the initial moments following the arrival of the laser pulse (hereinafter referred to as "state 1").

The electron density, derived from the narrow contour, falls within the range of $1.8\cdot10^{14}$-$5\cdot10^{15}$ cm$^{-3}$. This corresponds to the plasma density at later stages in the evolution of the hybrid plasma, generated by injecting laser power into the mw plasma, once the dense plasma clot has expanded and the electron density has reached quasi-stationary values. Hereafter, we will refer to this state of the hybrid plasma as "state 2." As the laser pulse energy increases, the electron density rises, reaching a maximum at an energy of approximately 33 mJ, before decreasing with any further increase in energy. The decline in electron density at $E_{pulse}$ >33 mJ is due to a reduction in both the laser pulse repetition rate and the average laser power.

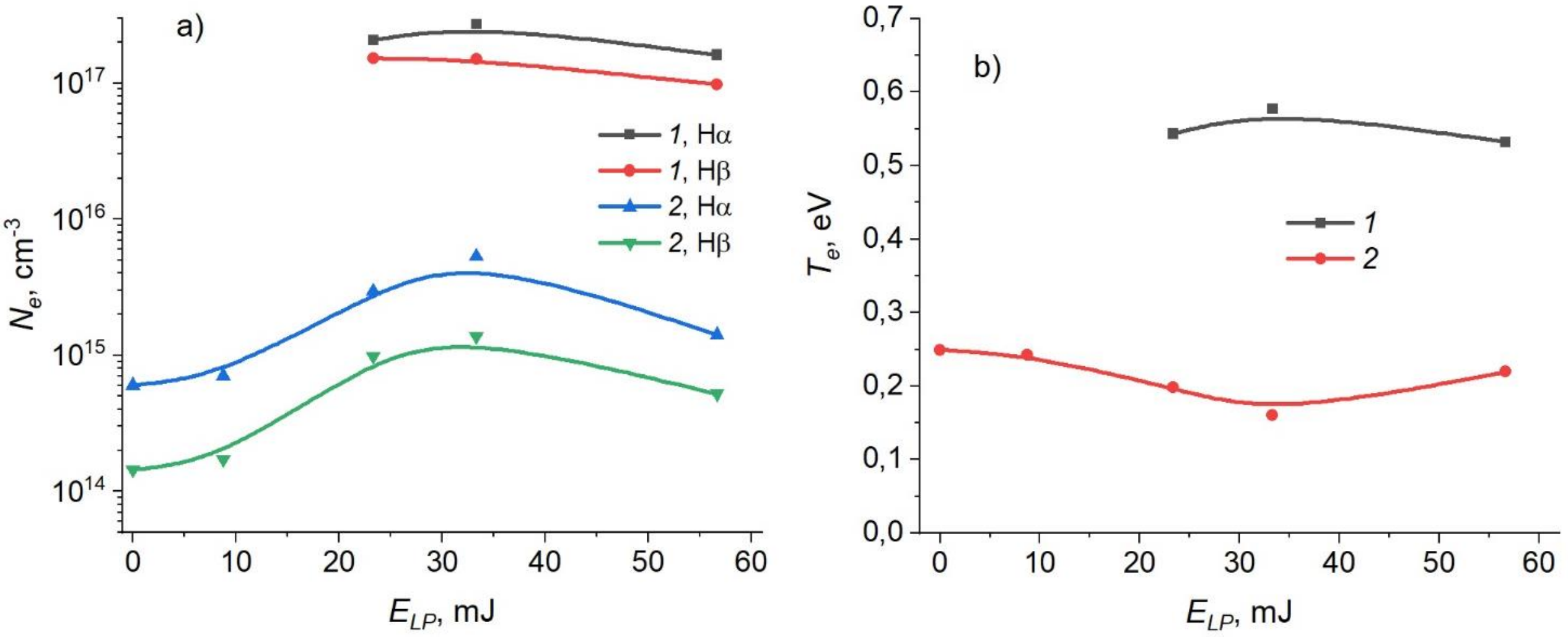

Fig. 5. Dependences of electron density (a) and temperature (b) values for two states of the hybrid plasma in the $95H_2{:}3Ar{:}2CH_4$ mixture on the energy of $CO_2$ laser pulses; *1* – state 1, *2* – state 2.

As can be seen from Figure 5a, the electron densities in state 1, calculated from the broad components of the dual-Lorentzian approximation of $H_\alpha$ and $H_\beta$ line profiles, differ only slightly. However, the electron densities in state 2, calculated from the narrow components of the same approximations, differ by a factor of approximately 3. This is primarily due to uncertainties in the dual-Lorentzian approximation of the $H_\alpha$ and $H_\beta$ line shapes, where the narrow component is reconstructed with a higher degree of error, particularly in the case of the relatively weak $H_\beta$ line.

Figure 5b shows electron temperatures calculated from the area ratio of the broad components, obtained via a dual-Lorentzian approximation of $H_\alpha$ and $H_\beta$ shapes, as well as from the area ratio of the narrow components derived from the same $H_\alpha$ and $H_\beta$ shapes. As can be seen from the figure, the electron temperature calculated from the area ratio of the broad components is on the order of 0.53–0.58 eV. The electron temperature calculated from the area ratio of the narrow components is on the order of 0.16–0.24 eV. Thus, in the initial moments following the arrival of the laser pulse, the electron temperature in the dense hybrid plasma region is 2–3 times higher than the electron temperature that is established after the dense plasma region has expanded.

The intensity of the $H_\alpha$ and $H_\beta$ lines corresponding to each of the plasma states is

$$I_{ki}^{j} = s(\lambda_{ki})\frac{hc}{\lambda_{ki}}A_{ki}N_n^{j}V_j\tau_P\nu_P, \quad (6)$$

where $s(\lambda_{ki})$ is sensitivity of the detection system, which depends on the wavelength, $\lambda_{ki}$ is spectral line wavelength, $A_{ki}$ is transition probability, $N_n^j$ is populations of hydrogen atom excited states, associated with hybrid plasma states ($n$=3, 4; $j$=1, 2), $V_j$ is plasma volume in state 1 or 2, $\tau_P$ is pulse duration, $\nu_P$ is pulse repetition frequency, $h$ and $c$ are known constants. Assuming that in each of the two plasma states the density of excited hydrogen atoms $H_n^j$ is proportional to the electron density $n_e^j$, and taking into account that the line emission intensity in each plasma state is proportional to the area of the corresponding line profile $I_{ki}^j \sim S_{H\alpha,\beta}^j$, we obtain a relationship for the volumes of the two states of the hybrid plasma:

$$\frac{V_1}{V_2} = \frac{S_{H\alpha,\beta}^1/(n_e^1 \cdot \tau_{LP} \cdot \nu_{LP})}{S_{H\alpha,\beta}^2/(n_e^2 \cdot \tau_{mw} \cdot \nu_{mw})}. \qquad (7)$$

Here, it is assumed that the duration of the dense plasma state is approximately equal to the duration of the $CO_2$ laser pulses, while the duration of the second state is equal to the duration of the mw pulse minus the duration of the laser pulse.

The volume ratios of two hybrid plasma states, calculated from the $H_\alpha$ and $H_\beta$ line profiles according to relation (7) as a function of the $CO_2$ laser pulse energy, are shown in Fig. 6.

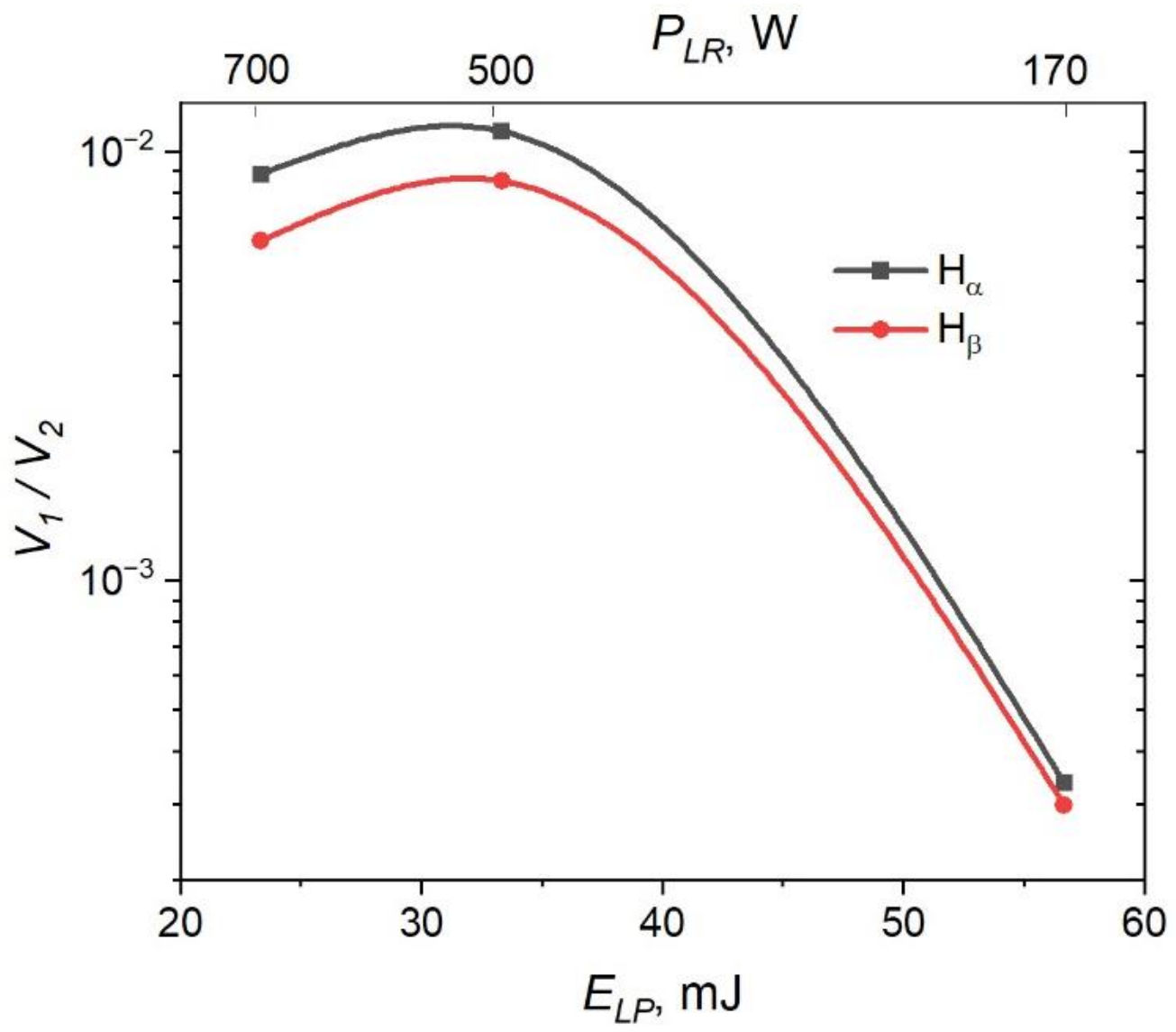

Fig. 6. Dependences of the characteristic volume ratio of two states of the hybrid plasma on the energy of $CO_2$ laser pulses; $V_1$ and $V_2$ are the volume of the hybrid plasma states 1 and 2.

The values obtained from the $H_\alpha$ line profiles are close to those derived from the $H_\beta$ line profiles. It can be seen that the volume of the dense plasma region reaches a maximum at a pulse energy of approximately 33 mJ (at a frequency of 15 kHz). Increasing the laser pulse energy while simultaneously decreasing the pulse frequency leads to a reduction in the volume of the dense plasma. If we conceptually represent the hybrid plasma as a cylinder with a volume $V_2 = \pi R^2 h$, and volume of dense plasma $V_1 = \pi r^2 h$ (here, $R$ is the plasma radius in state 2, which is approximately equal to the radius of the mw plasma; $r$ is the plasma radius in state 1; and $h$ is the height of the cylinder, which is equal to the diameter of the diagnostic window aperture), then for the ratio of the radii we obtain $\frac{r}{R} = (V_1/V_2)^{1/2}$. At laser radiation frequencies of 15 and 30 kHz, as follows from Fig. 6, $V_1/V_2 \sim 0{,}01$, and for the ratio of the radii, we obtain $\frac{r}{R}$~0.1. Assuming that the volume of state 2 of the hybrid plasma is approximately equal to the volume of the mw plasma, and given a characteristic radius of the mw plasma of ~5–10 mm, the radius of the dense plasma (state 1) is ~0.5–1 mm.

## Conclusion

The parameters of the electron component of $95H_2:3Ar:2CH_4$ atmospheric-pressure hybrid plasma sustained by pulsed microwave radiation (2.47 GHz) combined with $CO_2$ laser radiation (10.6 μm) were investigated using optical emission spectroscopy. It is shown that the $H_\alpha$ and $H_\beta$ atomic hydrogen lines in the hybrid plasma spectra are significantly broadened, and their line shapes can be described by the sum of two Lorentzian shapes (a dual-Lorentzian approximation).

To simplify the description of the hybrid plasma properties, a concept involving two plasma states, resulting from the pulsed injection of laser energy, has been introduced. Each state corresponds to one of the two contours used in a dual-Lorentzian approximation of the $H_\alpha$ and $H_\beta$ line shapes. Stark broadening of the $H_\alpha$ and $H_\beta$ lines was used to determine the electron density in the plasma, while the ratio of their intensities allowed for an estimation of the electron temperature. The broad contour corresponds to the plasma state during the laser pulse, characterized by a high electron density of $(1.0–2.5)\times10^{17}$ $cm^{-3}$ and an electron temperature of ~0.5–0.6 eV. The narrow contour corresponds to the hybrid plasma state during the intervals between laser pulses; in this state, the electron density is more than an order of magnitude lower (($1.8\times10^{14}$–$5\times10^{1}$ $cm^{-3}$)) than in the first state, yet 2–10 times higher than that of microwave (mw) plasma ($(1–5)\times10^{14}$ $cm^{-3}$). The electron temperature in this second state of the hybrid plasma is 0.16–0.24 eV.

Based on the area ratio of the broad and narrow contours of the $H_\alpha$ and $H_\beta$ line dual-Lorentzian approximation, the volume ratio of the two hybrid plasma states was estimated. For $CO_2$ laser emission frequencies of 15 and 30 kHz, a value of ~0.01 was obtained for the volume ratio $V_1/V_2$; this implies that the radius of the dense plasma region, characterized by an electron density of ~$10^{17}$ $cm^{-3}$, is approximately 0.5–1 mm.

## Funding

This work was supported by the Russian Science Foundation (RSF) grant No. 25-29-00816.

**Conflict of interest**

The authors declare that they have no conflict of interest.